\documentclass[
reprint, %reprint, preprint, reprint, or letter
superscriptaddress,
amsmath,
amssymb,
aps,
prl,
longbibliography,
]{revtex4-2}

\usepackage{graphicx}% Include figure files
\usepackage{dcolumn}% Align table columns on decimal point
\usepackage{physics}%for making Brakets
\usepackage{gensymb}% Degree symbol
\usepackage{bm}% bold math
\usepackage{xcolor}
\usepackage{footmisc}
\usepackage{xr}
\usepackage{textgreek}
\begin{document}

\preprint{APS/123-QED}

\title{Super-resolution Airy disk microscopy of individual color centers in diamond}% Force line breaks with \\
%\thanks{A footnote to the article title}%

\author{A. Gardill}
\author{I. Kemeny}
\author{Y. Li}
\affiliation{Department of Physics, University of Wisconsin, Madison, Wisconsin 53706, USA}

\author{M. Zahedian}
\affiliation{Department of Engineering Physics, University of Wisconsin, Madison, Wisconsin 53706, USA}

\author{M. C. Cambria}
\author{X. Xu}
\affiliation{Department of Physics, University of Wisconsin, Madison, Wisconsin 53706, USA}

\author{V. Lordi}
\affiliation{Lawrence Livermore National Laboratory, Livermore, CA, 94551, USA}

\author{\'A. Gali}
\affiliation{Wigner Research Centre for Physics, Institute for Solid State Physics and Optics, PO. Box 49, H-1525 Budapest, Hungary}
\affiliation{Department of Atomic Physics, Institute of Physics, Budapest University of Technology and Economics, M\H{u}egyetem rakpart 3., H-1111 Budapest, Hungary}

\author{J. R. Maze}
\affiliation{Instituto de F\'isica, Pontificia Universidad Cat\'olica de Chile, Casilla 306, Santiago, Chile}

\author{J. T. Choy}
\affiliation{Department of Engineering Physics, University of Wisconsin, Madison, Wisconsin 53706, USA}

\author{S. Kolkowitz}
\email{kolkowitz@wisc.edu}
\affiliation{Department of Physics, University of Wisconsin, Madison, Wisconsin 53706, USA}

\date{\today}% It is always \today, today,
             %  but any date may be explicitly specified
\begin{abstract}Super-resolution imaging techniques enable nanoscale microscopy in fields such as physics, biology, and chemistry. However, many super-resolution techniques require specialized optical components, such as a helical-phase mask. We present a novel technique, Super-resolution Airy disk Microscopy (SAM) that can be used in a standard confocal microscope without any specialized optics. We demonstrate this technique, in combination with ground state depletion, to image and control nitrogen-vacancy (NV) centers in bulk diamond below the diffraction limit. A greater than 14-fold improvement in resolution compared to the diffraction limit is achieved, corresponding to a spatial resolution of 16.9(8)~nm for a 1.3~NA microscope with 589~nm light. We make use of our enhanced spatial resolution to control the spins states of individual NV centers separated from each other by less than the diffraction limit, including pairs sharing the same orientation that are indistinguishable with a conventional electron spin resonance measurement.
\end{abstract}

\maketitle

\section{Introduction}

In recent years, super-resolution imaging techniques have pushed the achievable resolution of optical microscopy far below the diffraction limit \cite{hell2015sr_review, schermelleh2019sr_review, valli2021seeing}. Nanoscale imaging with visible light is now widely used in fields of biology \cite{willig2006}, chemistry \cite{cognet2008subdiffraction}, 
and physics \cite{bersin2019sr_nv_qi}. In quantum information applications, super-resolution techniques enable the localization and independent control of individual color centers at distances over which they can coherently interact \cite{Neumann2010, yao2012scalable, bersin2019sr_nv_qi}. Many current super-resolution techniques, however, either require stochastic processes and post-processing \cite{rust2006storm, betzig2006palm} or additional optics \cite{valli2021seeing, Klar2000}.
For example, stimulated emission depletion (STED) \cite{Hell1994, Klar2000} and ground state depletion (GSD) \cite{hell1995gsd, Bretschneider2007} techniques, used with fluorescent dyes \cite{Klar2000} and solid-state color centers \cite{ Hell2009, Wildanger2012, Treussart2013, Acosta2019}, rely on a doughnut-shaped beam profile to achieve super-resolution by deterministically switching off emitters outside of the field minimum at the doughnut center. However, the generation of the doughnut profile requires specialized optics such as helical-phase masks \cite{Wildanger2012, Treussart2013, Acosta2019}. The addition of these specialized optics either necessitates significant modifications to a pre-existing optical set up or a dedicated super-resolution microscope \cite{borlinghaus2016hyvolution}.

\begin{figure*}[htbp]
\includegraphics[width=\linewidth]{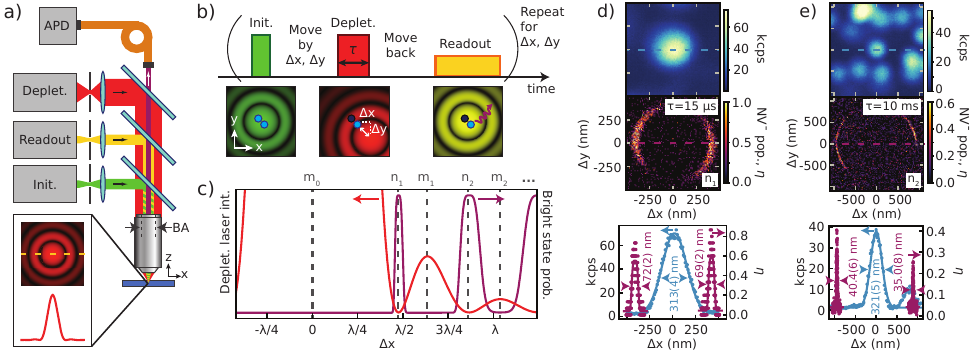}
\caption{\label{fig:fig1} Super-resolution Airy disk  Microscopy (SAM) technique. (a) Diagram of a confocal microscope. The finite back aperture (BA) of the objective produces an Airy disk at the focus, which is used for the SAM technique.
(b) Measurement sequence for the SAM technique. The position of the depletion pulse is scanned with respect to the emitter(s), while the initialization and readout pulses are applied centered on the emitter(s).
(c) Comparison of the intensity profile of the depletion pulse (red) and the expected probability of the optically bright state of the emitter (purple), demonstrating peaks at the at the Airy nodes ($n_1$, $n_2$, \dots) with widths below the diffraction limit. The Airy pattern intensity profile is calculated for the wavelength $\lambda$ of the depletion beam and an objective NA of 1.3 with no aberrations.
(d, e) Experimental data showing confocal (top) and SAM (middle) images of two different nitrogen-vacancy (NV) centers taken at two depletion pulse lengths. The bottom panel shows confocal (blue) and SAM (purple) line-cuts across the dashed lines, demonstrating resolution well below the optical diffraction limit for the SAM technique. 
}
\end{figure*}

\section{Experimental results}
In this paper, we introduce Super-resolution Airy disk Microscopy (SAM), a novel super-resolution technique that can be employed in a standard confocal microscope without requiring additional or specialized optics to isolate and localize single emitters below the diffraction limit. Figure~\ref{fig:fig1}(a) shows a diagram of a confocal microscope, where the finite size of the back aperture (BA) of a high numerical aperture (NA) microscope objective cuts off the outer radial components of the laser beam mode, resulting in the well-known Airy disk diffraction profile at the focus of the objective \cite{hechtoptics}. The central concept of the SAM technique is that the nodes, or dark rings, of the Airy disk pattern have close to zero laser intensity. As a result, for a sufficiently intense laser pulse, a two-level saturable emitter will saturate everywhere except in these rings. SAM can be employed with optical emitters that are compatible with STED \cite{Hell1994, Klar2000} or GSD microscopy \cite{hell1995gsd, Bretschneider2007}, but without requiring any of the additional optics needed to produce the doughnut-shaped beam typically used in these techniques. 

The sequence for the SAM technique, which is similar to the ``Spatial Photogeneration and Capture of chargE'' (SPaCE) sequence recently used with longer depletion pulses and at larger displacements to study charge dynamics in diamond \cite{Gardill2021}, is illustrated in Fig.~\ref{fig:fig1}(b). First, the initialization laser is focused at the location of one or more emitters, preparing the emitters into a desired optically bright ground state. Next, the depletion laser is focused at a distance $\Delta$x and $\Delta$y relative to the focus of the initialization pulse for time $\tau$. Under sufficiently high depletion laser intensity, the emitters will switch to an optically dark state unless one of the emitters lies in a ring of zero intensity of the Airy disk, in which case it will not be depleted. Finally, the readout laser is centered at the emitters and fluorescence is collected. This process is repeated for different depletion beam displacements ($\Delta$x, $\Delta$y) over a range on the order of 1~\textmu m, while the initialization and readout lasers are always centered on the emitter(s).

The simulated response of a single emitter to the SAM measurement scheme is shown in  Fig.~\ref{fig:fig1}(c), where the probability of the emitter being in the optically bright state is plotted (purple) as a function of the relative displacement of the depletion pulse, $\Delta$x. The Airy pattern of the depletion pulse is plotted (red) centered about $\Delta$x~=~0. Where the beam's intensity is high -- around the anti-nodes of the Airy pattern $m_0$, $m_1$, \dots -- the bright state of the emitter is depleted to the dark state. However, where the beam's intensity is lower than the bright state depletion threshold -- at the nodes $n_1$, $n_2$, $\dots$ -- the emitter remains in the bright state. For a sufficiently long depletion pulse $\tau$, the emitter will saturate around the nodes (first for $n_1$, then $n_2$ at longer $\tau$, and so on) and form a peak with a width no longer restricted by the diffraction limit.

We experimentally demonstrate SAM using the GSD technique \cite{hell1995gsd, Bretschneider2007} with single nitrogen-vacancy (NV) centers in an electronic grade diamond from Element Six. Specifically, we prepare an NV center in the NV$^-$ charge state with a 515~nm laser, deplete the NV center to the NV$^0$ charge state with a 638~nm laser, and measure the charge state with a 589~nm laser \cite{Shields2015} (experimental details in Supporting Information). Confocal scans of two different NV centers are shown in the top panels of Fig.~\ref{fig:fig1}(d) and (e). The resulting SAM measurements of the respective NV centers are shown in the middle panels for different depletion durations $\tau$. The SAM measurements plot the NV$^-$ state population, $\eta$, of the single NV center as a function of the depletion pulse's relative displacement ($\Delta$x, $\Delta$y) from the NV center. The resulting SAM measurements show a bright ring where the NV center remains in the NV$^-$ state. This ring corresponds to a displacement of the depletion pulse such that the node of the Airy pattern ($n_1$ or $n_2$) overlaps with the NV center and leaves it in NV$^-$, which fluoresces during the readout pulse. At other positions ($\Delta$x, $\Delta$y), the depletion beam converts the NV center to NV$^0$, which is not excited by the 589 nm readout pulse.

Figure~\ref{fig:fig1}(d) demonstrates the result of a relatively short depletion pulse, which produces a bright ring corresponding to the $n_1$ Airy disk node. A one-dimensional data set was collected along the line cuts for the confocal (blue) and SAM images (purple), which are directly compared in the bottom panel. The confocal data are fit to a Gaussian curve and show a full-width at half-max (FWHM) of 313(4)~nm. The peaks from the SAM measurement are individually fit to
\begin{equation}\label{quad_gaussian}
    \eta = A e^{-x^4/\sigma^4},
\end{equation}
where %$\eta$ is the NV$^-$ population, 
$A$ is the amplitude of the peak and $\sigma$ is related to the FWHM of the peak through ${\text{FWHM} = 2\sigma(\text{ln}(2) )^{1/4}}$. This fit corresponds to the SAM profile for a two-photon depletion process (as expected for the NV$^{-}$ center under illumination with 638~nm light \cite{Aslam2013}) responding to the intensity profile of the Airy pattern around the node (see Supporting Information for more details), and describes the observed peaks more accurately than a Gaussian fit. The data only take discrete values of the average charge state population due to the finite number of single-shot charge state measurements. The fit returns a FWHM of 69(2)~nm for the peak at $\Delta$x = +400~nm in the SAM measurement, representing a nearly 5-fold improvement in resolution over the confocal measurement, and a 3-fold improvement over the theoretical diffraction limit for a 1.3~NA microscope with 589~nm light, which represents the diffraction limit for the SAM technique with the exclusion of the depletion pulse.

Figure~\ref{fig:fig1}(e) shows a similar set of data for a second NV center using a longer pulse at which only the $n_2$ ring is visible, with peak FWHM of 35.0(8)~nm for the peak at $\Delta$x = +900~nm. This corresponds to a nearly 10-fold improvement in resolution from the confocal image, and nearly 7-fold improvement over the diffraction limit (1.3~NA, 589~nm light). The $n_2$ SAM rings are observed to reach narrower widths than the $n_1$ rings, which we attribute to the reduced intensity of $m_1$ and $m_2$ anti-nodes of the Airy disk relative to the intensity minimum of the node, and is consistent with our numerical simulations of the expected beam profile in the presence of aberrations (see Supporting Information).

\begin{figure}[b]
\includegraphics[width=\linewidth]{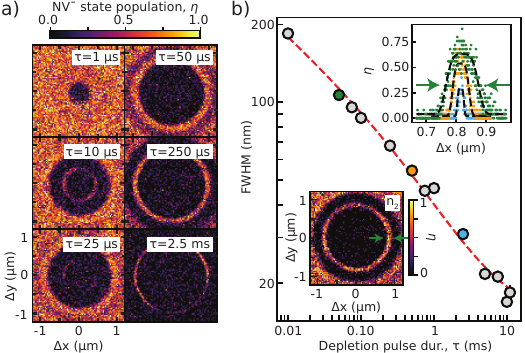}
\centering
\caption{\label{fig:fig2} SAM profile dependence on depletion pulse duration. 
(a) Evolution of the SAM super-resolution rings as a function of depletion pulse duration, $\tau$.
(b) FWHM of the $n_2$ super-resolution ring as a function of the depletion pulse duration, and fit (red dashed line, see Supporting Information). Error bars of one standard error are smaller than the data points.  The FWHM is extracted from data taken along the $\Delta$x direction; location highlighted with green arrows in the lower left inset. The upper right inset shows representative measured SAM profiles (colored points) fit to Eq.~\ref{quad_gaussian} (black dashed lines), corresponding to the respective colored data points in the main figure. 
}
\end{figure}

%\section{Results}
The evolution of the super-resolution rings of a single NV center with increasing $\tau$ is shown in Fig.~\ref{fig:fig2}(a). At 1~\textmu s, no rings are formed because the pulse is too short to deplete the NV center, except at center peak of the Airy pattern ($m_0$). With increasing $\tau$, the NV center is depleted by other parts of the beam, seen as the dark area growing radially outwards. However, the $n_1$ and $n_2$ rings initially remain bright. If the depletion pulse intensity at the nodes were zero, the rings at the nodes would continue to narrow indefinitely, however the $n_1$ ring disappears after 25~\textmu s and the $n_2$ ring disappears after 10~ms. The non-zero intensity at the nodes is most likely due to spherical aberrations in the depletion beam caused by the objective oil-diamond interface (see Supporting Information for more details). Additionally, the ring brightness is not uniform, with bright arcs lasting slightly longer that the rest of the ring, for example, in the $n_1$ ring at 25~\textmu s. We attribute this to the linear polarization of the depletion laser \cite{novotny2012nano_optics}, which causes higher intensity in the depletion beam along the direction of polarization. In our experiments, the depletion beam is linearly polarized at a small angle from the $y$-axis (see Supporting Information for more details). 

Just as with traditional STED or GSD microscopy there is no fundamental lower limit to the resolution of the SAM super-resolution ring \cite{Klar2000, willig2006}. Fig.~\ref{fig:fig2}(b) quantitatively evaluates the resolution of the SAM technique by taking a line-cut through the $n_2$ ring (taken along $\Delta$x in a line between the green arrows in the bottom left inset). Three  representative  measurements are shown in the top right inset and are fit with Eq.~\ref{quad_gaussian} to extract the FWHM. The corresponding values of FWHM are shown as a function of $\tau$ in Fig.~\ref{fig:fig2}(b) on a log-log plot. The resolution decreases with time down to a value of 16.9(8)~nm with a 10~ms pulse, after which the peak's height is below the noise floor. This demonstrates a 14-fold improvement in resolution compared to the theoretical diffraction limit in our microscope for the 589~nm light used for the readout pulse.  

The resolution dependence on $\tau$ in Fig.~\ref{fig:fig2}(b) can be understood by considering the effect of the Airy disk intensity pattern on the charge state dynamics of the NV center \cite{Aslam2013}, plotted in Fig.~\ref{fig:fig1}(c). Two additional assumptions must be made to accurately describe the experimental results: 1) the intensity of light of the depletion pulse at the $n_2$ node is not zero, but is some small fraction, $\epsilon$, of the maximum intensity of the beam, $I_0$ and 2) there is some technical imprecision in repeatedly positioning the beam with respect to the NV center, $\Delta R$, setting a lower bound on the resolution. The data is fit to the expected resolution scaling (see Supporting Information for details) and plotted as the red dashed line in Fig.~\ref{fig:fig2}(b). The resolution also depends on the depletion laser power (see Supporting Information), and offers a trade-off with pulse duration -- lower depletion power can be used with longer pulse durations to achieve the same resolution. From the fit, we extract a background intensity at $n_2$ of $\epsilon=0.44(12)\%$ of $I_0$, compared to typical values of $\epsilon=1\%-5\%$ of $I_0$ for doughnut beam profiles created for STED techniques \cite{neupane2013,Acosta2019,  Hell2021}.  We find a lower bound to our resolution of $\Delta R$ = 14(3)~nm, likely limited by a combination of thermal drifts of the sample relative to the microscope and the repeatability of the piezo stage used for these experiments, which is specified as $\pm5$~nm. 

\begin{figure}[t]
\includegraphics[width=\linewidth]{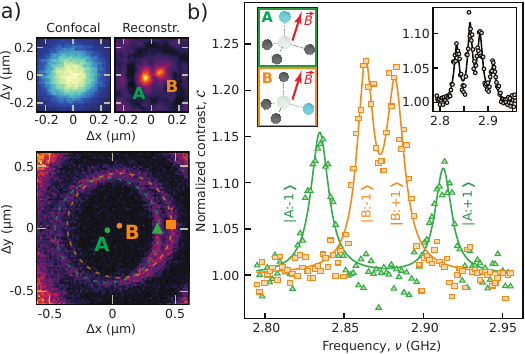}
\centering
\caption{\label{fig:fig3} SAM electron spin resonance (ESR).
(a) Confocal (top left) and SAM measurement (bottom) of the same pair of NV centers NV$_\text{A}$ and NV$_\text{B}$ (their positions marked by a green and orange dot, respectively). 
%An image of the NV centers is visually reconstructed (top right) by plotting the cost function used to fit to the SAM data.
A super-resolution image of the NV centers is reconstructed (top right) by plotting the cost function used to fit to the super-resolution rings in the SAM measurement (see Supporting Information).
(b) ESR measurement using spin-to-charge readout and performed with the SAM technique to isolate NV$_\text{A}$ (green triangles) or NV$_\text{B}$ (orange squares), with corresponding fits. The left inset depicts the NV centers' orientations in the diamond lattice with respect to the applied magnetic field $\vec{B}$. The right inset shows a standard ESR measurement at the same magnetic field without the SAM technique.
}
\end{figure}

The SAM technique can be used to localize and independently manipulate the electronic spin states of pairs of NV centers separated by less than the diffraction limit. Figure~\ref{fig:fig3}(a) shows a confocal scan (top left) of an unresolved pair of NV centers NV$_\text{A}$ and NV$_\text{B}$, as well as a corresponding SAM measurement (bottom) revealing two super-resolution rings. A circle fitting routine is used to identify the centers of the super-resolution rings in the SAM data. %fit rings to the SAM data to determine the rings' centers -- 
Each ring center corresponds to the position of the NV center that generates the ring (see Supporting Information). 
%respective NV centers' positions (see Supplemental Information for more details). 
A super-resolution image of the NV pair can be reconstructed (top right) from the SAM data by plotting the value of the cost function used in the circle fitting routine (Supporting Information) 
%(not obvious why cost function reflects the superresolution experiment and can be used to describe the data) 
%used to fit a circle to the super resolution rings (top right) 
at each position $x$ and $y$ for an optimized circle radius. 
We note that the super-resolution reconstruction images shown here are produced from SAM measurements
%We note that this image was reconstructed from the SAM data shown in the lower panel, 
with fixed initialization and readout pulses positioned at the image origin, and the reconstructed image range is limited to roughly the size of the confocal spot. To produce an image with a larger field of view with more emitters, the initialization and readout pulse locations could be rastered across the sample, with the measurement protocol demonstrated here repeated at each location.
%$(0,0)$ in the confocal images To produce a larger image of a larger field of view with more emitters, the SAM measurement could be rastered over a larger area.  %Lower values indicate positions in the $xy$-plane where there is more likely to be an NV center.

From the center of the fitted rings in Fig.~\ref{fig:fig3}(a), we find the two NV centers are separated by 113(26)~nm in the $xy$-plane -- below the diffraction limit of our confocal microscope for visible light. We note that the SAM technique does not significantly increase the resolution in $z$, as with many other super-resolution techniques such as STED and GSD \cite{hell2015sr_review, schermelleh2019sr_review}. NV$_\text{A}$ (NV$_\text{B}$) is isolated with a SAM electron spin resonance measurement by applying a depletion pulse at the green triangle (orange square) in Fig.~\ref{fig:fig3}(a), so that the $n_1$ node of the Airy disk falls on NV$_\text{A}$ (NV$_\text{B}$) while NV$_\text{B}$ (NV$_\text{A}$) is converted to NV$^0$. Next, a microwave pulse is applied at frequency $\nu$ and the spin state of NV$_\text{A}$ (NV$_\text{B}$) is measured using a spin-to-charge readout technique \cite{Shields2015} (see Supporting Information for details on experimental procedure).

Figure~\ref{fig:fig3}(b) shows the SAM ESR measurement on either NV$_\text{A}$ (green triangles) or NV$_\text{B}$  (orange squares). An external magnetic field of roughly 30~G is applied with a permanent magnet to split the $\ket{+1}$ and $\ket{-1}$ ground state spin states. The two NV centers have different orientations within the diamond (left inset), and experience different projections of the magnetic field on their axes, causing the spin states to split by different amounts \cite{Doherty2013}. The SAM ESR measurement of NV$_\text{A}$ exhibits no peaks from $\ket{\text{B}:\pm1}$ and vice verse. For comparison, the right inset shows a standard ESR measurement without application of the SAM depletion pulse, in which the resonances of both NV centers are clearly present.

\begin{figure}[htbp]
\includegraphics[width=\linewidth]{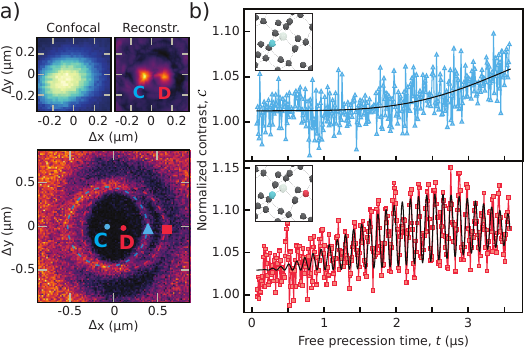}
\centering
\caption{\label{fig:fig4} SAM spin echo.
Confocal (top left) and SAM measurement (bottom) of the same pair of NV centers NV$_\text{C}$ and NV$_\text{D}$ (their positions marked by a blue and red dot, respectively). 
%Note that other NV centers are present in this cluster, giving the spot in the confocal image an asymmetric shape; 
The asymmetric shape of the feature in the confocal image indicates the presence of at least one additional nearby NV center; however, the position of the readout beam in the SAM measurements isolates NV$_\text{C}$ and NV$_\text{D}$. 
%The positions of these two NV centers are visually reconstructed (top right).
A super-resolution reconstruction image of the two NV centers is shown at top right. 
(b) Spin echo measurements \cite{childress2006spinecho} using spin-to-charge readout and performed with the SAM technique to isolate NV$_\text{C}$ (blue triangles) and NV$_\text{D}$ (red squares), along with fits (black solid lines, see Supporting Information). The insets depict the corresponding local nuclear spin environments of the two NV centers, with a strongly coupled  $^{13}$C atom nearby NV$_\text{D}$ and none for NV$_\text{C}$ \cite{Smeltzer2011}.
}
\end{figure}

The SAM technique can also be applied to resolve closely spaced NV centers with the same orientation and to determine their unique respective nuclear spin environments. A confocal image (top left) of a cluster of NV centers is shown in Fig.~\ref{fig:fig4}(a). 
%The cluster is elongated along one direction, meaning that not all the NV centers of this cluster are captured in the confocal spot size. This allows us to isolate just two NV centers, NV$_\text{C}$ and NV$_\text{D}$, in the SAM measurements by fixing the readout pulse position to the origin. 
%The asymmetric shape of the cluster indicates that it contains 3 or more NV centers
The NV cluster displays reflection symmetry about only a single axis, indicating that it must consist of at least three NV centers. By positioning the readout pulse at $(0,0)$, we collect fluorescence primarily from just two NV centers: NV$_\text{C}$ and NV$_\text{D}$, with the third NV center contributing little to no light.
%The cluster displays reflection symmetry about only a single axis (which axis? indicate this in the fig), indicating that it must consist of at least three NV centers. 
%At least one of these NV centers is farther from the center of the cluster, which allows us to isolate just two NV centers, NV$_\text{C}$ and NV$_\text{D}$, in the SAM measurements. 
%(A little confusing. NVc and NVd are the two NVs from which the one is further away right? so I might say something like- At least one of these NV centers is farther from the center of the cluster, which allows us to isolate the other two NVs, NVC and NVD in the SAM measurements). 
A SAM measurement (bottom) and corresponding super-resolution reconstruction image of the two NV centers (top right) 
%and the corresponding SAM image (bottom) 
are also shown in Fig.~\ref{fig:fig4}(a). The former resolves two individual NV centers, identified by distinct rings with centers marked as blue and red dots for NV$_\text{C}$ and NV$_\text{D}$, respectively. They share the same orientation (see Supporting Information) and are found to be 181(26)~nm apart in the $xy$-plane. We use SAM to apply super-resolved spin echo \cite{childress2006spinecho} microwave pulse sequences to both NV centers individually. The 8.870$\pm$0.003~MHz oscillations apparent in the spin echo signal of NV$_\text{D}$ indicate a strongly coupled $^{13}$C on a nearby lattice site, which is absent for NV$_\text{C}$. The value of the coupling strength indicates that the $^{13}$C atom is located two lattice sites from NV$_\text{D}$'s vacancy \cite{Smeltzer2011}, as depicted in the lower inset.

\section{Conclusion and outlook}
The SAM technique is a novel super-resolution technique that offers nanometer spatial resolution with a basic confocal microscope using off the shelf components. Without modifications to our confocal microscope, we achieved a 14-fold improvement in resolution over the diffraction limit. The minimum resolution we achieved is likely limited by the repeatability of our positioning hardware and thermal drifts of our apparatus, and could be further improved. In addition, the resolution could potentially be improved by reducing spherical aberrations.

We also demonstrated the versatility of the SAM technique by combining it with coherent spin manipulation of NV centers. We performed individual ESR and spin echo measurements on NV centers isolated from pairs of centers separated by less than the diffraction limit. While the super-resolution images shown in Figs.~\ref{fig:fig3} and \ref{fig:fig4} are reconstructed from SAM measurements with fixed initialization and readout pulse positions,
%While we show reconstructed visual images of the NV centers positions in Figs.~\ref{fig:fig3} and \ref{fig:fig4}, 
the SAM technique could be extended to larger fields of view by scanning the spatial position of the readout and initialization pulses and repeating the SAM measurement protocol at each location. %, in which the initialization and readout beams would be moved in $x$ and $y$.
%to larger fields of view by also scanning the initialization and readout beams.
%a more conventional imaging approach by co-scanning the initialization, depletion, and readout beams -- with the spatial position of the depletion beam offset from the other two beams -- which would easily lend itself to emitters with long-lived depletion states for which the laser pulses could be applied separately in time. 
%by scanning both the readout beam and the depletion beam, such that the location of the depletion beam can be independently displaced with respect to the location of the readout beam. 
Finally, we note that the SAM technique is not specific to NV centers; it could be applied to other fluorescent emitters used in STED or GSD \cite{Acosta2019, Hoyer_Quantum_Dot_SR, Kolesov_YAG_SR} provided that either the depleted state is long-lived or that the depletion beam can be positioned independently from the readout beam so that they can be applied simultaneously.

\section{Funding}
\vspace{0.5cm}
\noindent Experimental work, data analysis, and theoretical efforts conducted at UW--Madison were supported by the U.S. Department of Energy, Office of Science, Basic Energy Sciences under Award \#DE-SC0020313. Part of this work by V.L.~was performed under the auspices of the U.S. Department of Energy at Lawrence Livermore National Laboratory under Contract DE-AC52-07NA27344.  \'A.G.~ acknowledges the Hungarian NKFIH grant No.\ KKP129866 of the National Excellence Program of Quantum-coherent materials project and the support for the Quantum Information National Laboratory from the Ministry of Innovation and Technology of Hungary, and the EU H2020 Quantum Technology Flagship project ASTERIQS (Grant No.\ 820394). J.R.M.~acknowledges support from ANID-Fondecyt 1180673 and ANID-Anillo ACT192023. A.G.~acknowledges support from the Department of Defense through the National Defense Science and Engineering Graduate Fellowship (NDSEG) program.

\section{Acknowledgements}
\vspace{0.5cm}
\noindent The authors thank Jeff Thompson, Hossein Dinani, and Ariel Norambuena for enlightening discussions and helpful insights on the manuscript. 

\section{Disclosures.} 
\vspace{0.5cm}
\noindent The authors declare no financial or commercial conflicts of interest.

%\section{Data availability.}
%\vspace{0.5cm}
%\noindent The main data supporting the results in this study are available within the main text and Supplement 1. Raw datasets are available for research purposes from the corresponding authors upon reasonable request.

\bibliography{MainReferences}

\end{document}

\preprint{APS/123-QED}

\title{Supplemental Information for ``Super-resolution Airy disk microscopy of individual color centers in diamond''}

\author{A. Gardill}
\author{I. Kemeny}
\author{Y. Li}
\affiliation{Department of Physics, University of Wisconsin, Madison, Wisconsin 53706, USA}

\author{M. Zahedian}
\affiliation{Department of Engineering, University of Wisconsin, Madison, Wisconsin 53706, USA}

\author{M. C. Cambria}
\author{X. Xu}
\affiliation{Department of Physics, University of Wisconsin, Madison, Wisconsin 53706, USA}

\author{V. Lordi}
\affiliation{Lawrence Livermore National Laboratory, Livermore, CA, 94551, USA}

\author{\'A. Gali}
\affiliation{Wigner Research Centre for Physics, Institute for Solid State Physics and Optics, PO. Box 49, H-1525 Budapest, Hungary}
\affiliation{Department of Atomic Physics, Institute of Physics, Budapest University of Technology and Economics, M\H{u}egyetem rakpart 3., H-1111 Budapest, Hungary}

\author{J. R. Maze}
\affiliation{Instituto de F\'isica, Pontificia Universidad Cat\'olica de Chile, Casilla 306, Santiago, Chile}

\author{J. Choy}
\affiliation{Department of Engineering, University of Wisconsin, Madison, Wisconsin 53706, USA}

\author{S. Kolkowitz}
\email{kolkowitz@wisc.edu}
\affiliation{Department of Physics, University of Wisconsin, Madison, Wisconsin 53706, USA}

\maketitle

\section{Experimental Details}

\subsection{SAM measurements}
Here we provide the measurement parameters (laser power and pulse duration) used to perform Super-resolution Airy disk Microscopy (SAM) on nitrogen vacancy (NV) centers as depicted in measurement sequence Fig.~1(b) of the main text. The initialization pulse is 1~mW at 515~nm for 10~\textmu s. The depletion pulse is 20~mW at 638~nm for a variable duration. The readout pulse is 10~\textmu W at 589~nm for 50~ms. 

In the single NV center SAM experiments, we perform single-shot readout of the NV center's charge state, which is assumed to be either NV$^0$ or NV$^-$. Under 589~nm readout illumination, the fluorescence of NV$^0$ is less than that of NV$^-$, allowing the charge state of the NV center to be assigned after a single measurement based on whether the recorded number of counts is above or below a predetermined threshold \cite{Shields2015}. For the experiments in the main text and the Supplemental Information, our single-shot readout fidelity is approximately 85$\%$. The population data shown in the one-dimensional (two-dimensional) SAM experiments in the main text consist of $\sim 100$ ($\sim 5$) averaged single-shot readout measurements per spatial point.

The SAM technique relies on the ability to repeatably move the sample relative to the beam's focus in the focal plane of the objective. In this paper, data is presented from two separate confocal microscopes that accomplish this in two distinct ways.  Confocal microscope A (CFM A) uses a closed-loop two-axis piezo nano-positioner to move the sample while confocal microscope B (CFM B) uses a scanning two-mirror galvonometer to move the beam's focus. CFM B was used for the spin measurements presented in Figs. 3 and 4 of the main text. CFM A was used for measurements presented in Figs. 1 and 2 of the main text and data presented in the Supplemental Information.  The nano-positioner in CFM A is quoted to have positioning repeatability of $\pm$5~nm. We note that the resolution of SAM measurements could be improved by reducing the positioning repeatability of the positioning hardware. Both CFM A and CFM B used the same 1.3~NA oil objective. 

\subsection{SAM - ESR and spin echo}

\begin{figure}[htbp]
\includegraphics[width=\textwidth]{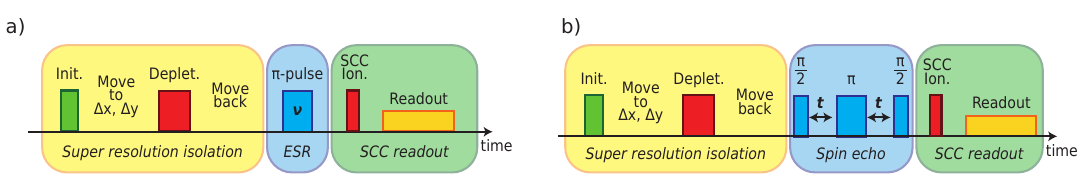}
\centering
\caption{\label{fig:ESR_echo} SAM electron spin resonance (ESR) and spin echo sequences.
(a) Measurement sequence for ESR measurements using the SAM technique. The target NV center is isolated by applying the depletion pulse ($\tau = 15$~\textmu s) at relative position ($\Delta$x, $\Delta$y), depleting the other NV center but leaving the target in NV$^-$. After isolation, a $\pi$-pulse of variable frequency $\nu$ is applied, and finally spin-to-charge readout is performed, with the spin-to-charge conversion pulse applied at the same location as the initialization pulse.
(b) Measurement sequence for spin echo measurements using the SAM technique.  After isolation with a depletion pulse of $\tau = 10$~\textmu s, a spin echo microwave measurement is performed with variable wait time $t$, and the spin state is read out with spin-to-charge conversion.
}
\end{figure}

Figure 3 of the main text presents electron spin resonance (ESR) measurements using the SAM technique. The measurement sequence is shown in Fig.~\ref{fig:ESR_echo}(a), which can be applied to a pair of NV centers that are not separable with standard confocal microscopy. First, the SAM technique is applied, starting with an initialization pulse followed by a depletion pulse ($\tau = 15$~\textmu s) at position ($\Delta$x, $\Delta$y), such that one of the NV centers falls within the node of the Airy pattern and the other does not, selectively ionizing the latter NV. For the rest of the measurement, the depleted NV center remains in the dark NV$^0$ charge state. A microwave $\pi -$pulse of varying frequency $\nu$ is applied. If $\nu$ is on resonance with the transition, the  $\pi -$pulse will drive the $m_s$=0 to $m_s$=$\pm$1 transition of the non-depleted NV center.  Lastly, the spin state of the non-depleted NV center is measured with a spin-to-charge readout technique \cite{Shields2015, Hopper2016scc, jayakumar2018scc}, where a short 300 ns ionization pulse (638~nm, 20~mW) is used to selectively ionize the non-depleted NV center if its spin state is $m_s$=0. This is followed by a 589~nm charge state readout pulse (2~ms, 15~\textmu W). This sequence is repeated without the $\pi$-pulse to record a reference for the counts in the $m_s=0$ state, which is then used to normalize the recorded counts. In contrast to the fluorescence dips observed with standard spin readout using green illumination, the normalized counts will be higher when $\nu$ is on resonance for spin-to-charge readout because the $m_s$=0 spin state is selectively ionized.

Figure 4 of the main text presents spin echo measurements  using the SAM technique. The measurement sequence is shown in Fig.~\ref{fig:ESR_echo}(b), which is nearly identical to the ESR measurement in (a), except in this case the microwave sequence is a spin echo pulse sequence \cite{childress2006spinecho, maze2008magneticsenseing}. Again, the data is normalized by recording the counts in the $m_s=0$ state. The basic pattern of SAM isolation followed by NV spin experiment and finally spin-to-charge readout can be adapted to work with any spin experiment by tailoring the microwave pulse sequence.

The data in Fig. 4 of the main text is fit to the following expected functional form for a spin echo signal from an NV center with a strongly coupled ${}^{13}$C nuclear spin \cite{childress2006spinecho}:
\begin{equation}\label{spin_echo}
  \mathcal{C} = \mathcal{C}_0 - e^{-(t / t_C)^4} (a - b \sin^2(2\pi f_0 t / 2) \sin^2(2\pi f_1 t / 2)),
\end{equation}
where $\mathcal{C}_0$ is the contrast, $a$ is the decay amplitude, $t_C$ is the coherence time, $f_{0,1}$ are oscillation frequencies for the nuclear spin due to its hyperfine coupling to the NV electronic spin, and $b$ is the oscillation amplitude. For data from NV$_\text{C}$, the oscillation amplitude, $b$, is set to zero because there is no sign of oscillations in the signal. From the the fit for NV$_\text{D}$, we extract an oscillation frequency of 8.870$\pm$0.003~MHz, corresponding to the hyperfine coupling strength between the NV electron spin and the nearby ${}^{13}$C spin. 

Lastly, we point out the time constraint imposed on the ``super-resolution isolation" portion (Fig.~\ref{fig:ESR_echo}) of these experiments due to the NV center's relaxation time. After the initialization pulse polarizes the NV center into $m_s = 0$, the NV center's spin naturally relaxes out of the $m_s = 0$ spin state with a typical relaxation time of $T_1 \sim5$~ms \cite{Jarmola2012, bar_gill2013}. This limits how long the depletion pulse and the movement to and from the beam's displacement ($\Delta$x, $\Delta$y) can take before the spin completely relaxes. In our SAM spin measurements the depletion pulse is short compared to $T_1$, however it takes 0.5~ms to move the focus each way, thus $\sim 1$ ms of spin relaxation elapses during the SAM sequence. This results in a reduction of spin contrast of $e^{-1/5} \sim 80\%$ in our measurements.  

\subsection{ESR data for NV$_\text{C}$ and NV$_\text{D}$}

\begin{figure}[htbp]
\includegraphics[width=\textwidth]{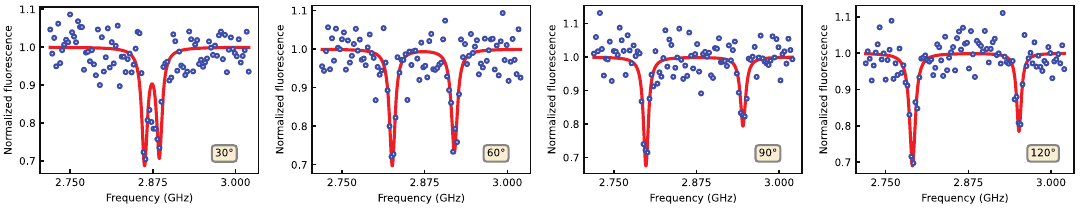}
\centering
\caption{\label{fig:nvc_d_esr} Standard ESR measurements on NV$_\text{C}$ and NV$_\text{D}$ for different orientations of an applied D.C. magnetic field. Only a single pair of resonances is seen, meaning that NV$_\text{C}$ and NV$_\text{D}$ must share the same orientation in the diamond.}
\end{figure}

The two NV centers presented in Fig. 4 of the main text share the same orientation in the diamond. This was verified by performing ESR measurements using standard 515~nm readout on the NV pair for different orientations of an applied D.C. magnetic field. The NV center's ground state spin levels $m_s = \pm1$ experience Zeeman splitting from the projection of the magnetic field on the NV center's axis (Fig.~3(b) inset) \cite{Doherty2013}, which is observed in an ESR measurement by the splitting between the resonant dips. Fig.~\ref{fig:nvc_d_esr} shows only one pair of resonant dips from these two NV centers regardless of the orientation of the magnetic field, indicating that they share the same orientation.% do not have different orientations. 

\subsection{Fourier transform of spin echo data}

\begin{figure}[htbp]
\includegraphics[width=1\textwidth]{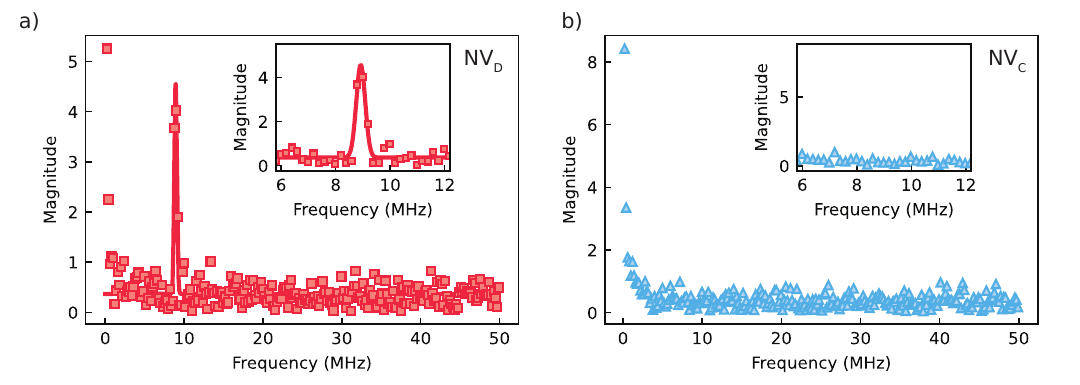}
\centering
\caption{\label{fig:fourier} Fourier transform of spin echo data from main text. (a) Fourier transform of NV$_\text{D}$ spin echo (red squares). Inset zooms in on data between 6 and 12~MHz, which is fit with Gaussian curve (red line). (b) Fourier transform of NV$_\text{C}$ spin echo (blue triangles). Inset zooms in on data between 6 and 12~MHz.
}
\end{figure}

The spin echo SAM measurement in Fig. 4(b) of the main text for NV$_\text{D}$ shows strong oscillations. In the main text, a fit to the data gives an oscillation frequency of 8.870$\pm$0.003~MHz. Figure \ref{fig:fourier}(a) shows the Fourier transform of the NV$_\text{D}$ spin echo data from 0 to 5~\textmu s. The inset highlights a single peak around 9~MHz. The peak is fit to a Gaussian, yielding an oscillation frequency of 8.92 $\pm$ 0.01~MHz. Figure \ref{fig:fourier}(b) shows the Fourier transform of the spin echo data of NV$_\text{C}$ with no identifiable peaks. This indicates that NV$_\text{C}$ has no strongly coupled nuclear spin, and also that the SAM spin echo signal obtained for NV$_\text{C}$ has no contribution from NV$_\text{D}$, confirming that the SAM technique is capable of isolating nearby NV centers with the same orientation for independent coherent spin measurements.

\subsection{Resolution as a function of power}

\begin{figure}[tbp]
\includegraphics[width=0.4\textwidth]{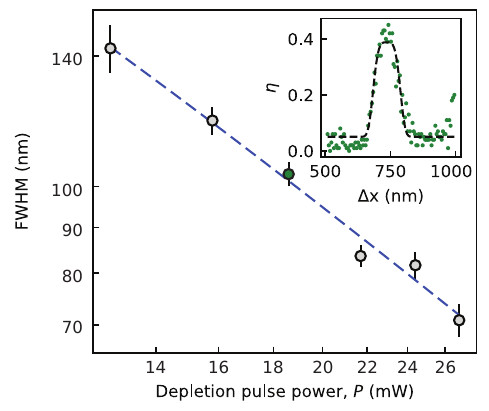}
\centering
\caption{\label{fig:power_dep} SAM profile dependence on depletion pulse power. The FWHM (grey circles) is fit as a function of depletion pulse power to  Eq.~\ref{background_int_width_scaling} (blue dashed line). The inset shows a representative one-dimensional SAM measurement (green points) across the $n_2$ ring. The FWHM is extracted using Eq.~\ref{eta_approx} and corresponds to the colored data point in the main figure.
}
\end{figure}

The SAM measurements presented in the main text used 20~mW depletion pulse. Here we show the effects of varying the depletion power on the SAM measurements. The dependence of the resolution with the depletion pulse power is shown in Fig.~\ref{fig:power_dep}, which is found by fitting one-dimensional data (an example of which is shown in the inset) to Eq.~\ref{eta_approx}, where we assume the depletion pulse power $P\propto I_0$. The FWHM data is then fit to the expected scaling of the resolution with power, Eq.~\ref{background_int_width_scaling}, by fixing the depletion duration $\tau$.

\subsection{Verifying scale of images in confocal microscopes}

\begin{figure}[btp]
\includegraphics[width=1\textwidth]{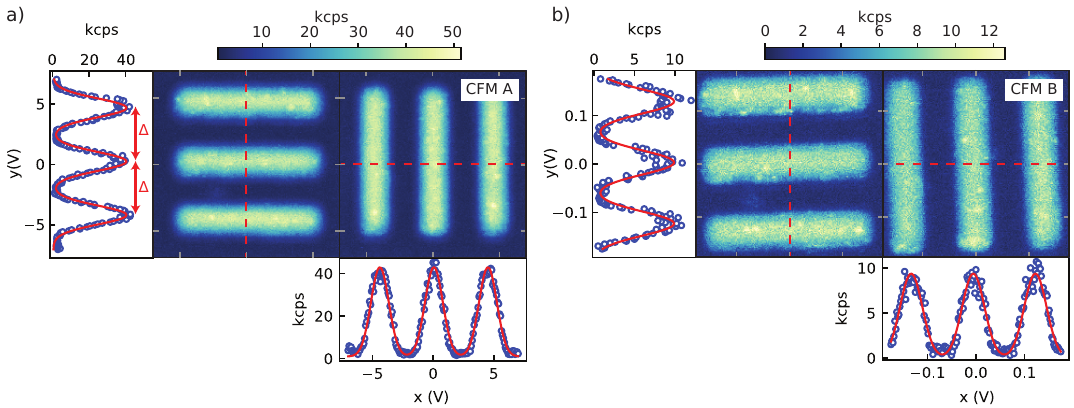}
\centering
\caption{\label{fig:usaf1951} Confocal images of resolution test target. (a) (b) Confocal images in CFM A (CFM B) of line pairs in elements 6 of group 7 of 1951 USAF resolution test target used to calibrate physical dimensions of measurements in CFM A (CFM B). Side and bottom panels plot line cuts of confocal data taken along dashed red line. A periodic sum of identical Gaussian curves (red solid line) is fit to the data, with periodic spacing $\Delta$. Both images were made with 515~nm illumination.
}
\end{figure}

Spatial displacements for all confocal and SAM measurements were calibrated using a 1951 US Air Force resolution test target (Thorlabs Negative 1951 USAF resolution test target, Ø1), which has vacuum-sputtered chrome patterns on a glass substrate. We use a negative test target where the patterns consist of voids in the chrome covering the remainder of the substrate. The patterns consist of three vertical and horizontal lines with standardized dimensions. Fig.~\ref{fig:usaf1951} (a) and (b) show confocal scans of line elements 6 of group 7, which are the smallest horizontal and vertical line pairs on this test target. The width of one line is 2.19~\textmu m. Fig.~\ref{fig:usaf1951}(a) was recorded with CFM A, while Fig.~\ref{fig:usaf1951}(b) was recorded with CFM B. The x- and y-axes of the scans show the voltages applied to the positioning equipment at each position to shift the location of the confocal microscope's focal spot on the test target. Also note, a higher laser power was used to take scans with CFM A than with CFM B.

The side and bottom panels in Fig.~\ref{fig:usaf1951}(a-b) show representative vertical and horizontal line cuts from the dashed red lines in the confocal images. The profile is well described by a fit to a periodic sum of identical Gaussians (red solid lines). In order to calibrate the ratio of applied voltage to focal spot shift, we compare the periodic spacing between these Gaussians, $\Delta$, to the corresponding standardized length of 4.38~\textmu m. We average this over three line cuts taken at equally separated positions from the respective confocal scan. Note that while the image for CFM B is tilted, we estimate this tilt to be $< 2 \degree$, corresponding to a calibration error of 0.03$\%$, which is smaller than any positional uncertainty quoted in the paper.

\section{Theoretical details}
\subsection{NV charge state response to Airy pattern}

The response of the NV center to the depletion pulse can be described by considering the intensity function of an Airy pattern and the charge state dynamics of the NV center. In this section we neglect the effects of aberrations, but we consider these effects using numerical simulations in a later section of this supplement.

An Airy pattern is formed when a plane wave is apertured by a lens of radius $a$ with a focal length of $L$. Assuming that the wavelength of light, $\lambda$ is much smaller than $L$, the ideal diffraction pattern formed at the focal plane is, in polar coordinates \cite{hechtoptics},
\begin{equation}\label{Airy}
     I(x,\phi) = I(x) = I_0 ( 2 J_1(x) / x)^2,
\end{equation}
where \(\phi\) is the angle about the optical axis, $I$ is the intensity of the pattern, $I_0$ is the maximum intensity of the pattern, $J_1$ is the first order Bessel function of the first kind, and $x$ is a dimensionless radial distance from the center of the pattern. The parameter $x$ is defined $x = \frac{2 \pi N r}{\lambda}$, where $N$ is the numerical aperture of the lens and $r$ is the radial distance from the center of the pattern.

Under 638~nm light, the NV$^-$ charge state converts to the NV$^0$ state through a two-photon process described by the rate $g_1$ \cite{Aslam2013}
\begin{equation}\label{g1}
g_1 = \nu_1  I(x)^2,
\end{equation}
where $\nu_1$ is a constant of proportionality and $I(x)$ is the intensity of the 638~nm light at radial distance $x$. We assume that 638~nm light negligibly drives transitions of NV$^0$ back to NV$^-$ \cite{Aslam2013}.

For an NV center that is initially in the NV$^-$ charge state, the population of NV$^-$, $\eta$, under 638~nm light applied for duration $\tau$ is given by
\begin{equation}\label{NV-_rate}
\eta = e^{-g_1 \tau} = e^{-\nu_1 I(x)^2 \tau}.
\end{equation}

Using Eq.~\ref{Airy} for the functional form of the intensity, the NV center's response to the Airy pattern is
\begin{equation}\label{eta}
\eta = e^{-\nu_1  I_0^2 ( 2J_1(x) / x)^4\tau}.
\end{equation}
This equation is plotted as the purple line in Fig. 1(c) of the main text.

Next we consider the approximate form of Eq.~\ref{eta} for $x$ near a node of the Airy pattern ($n_1, n_2, \dots$). We expand Eq.~\ref{Airy} about $x = x_1$, where $x_1$ is the position of the $n_1$ node. To leading order,
\begin{align}
I(x)/I_0 &= (2(J_0(x_1) - J_2(x_1)) / x_1)^2 (x - x_1)^2 + (\dots)\nonumber\\ 
&\approx c (x - x_1)^2,
\end{align}
where the coefficient \(c=(2(J_0(x_1) - J_2(x_1)) / x_1)^2\). Thus the NV$^{-}$ population around the Airy nodes obeys
\begin{equation}\label{eta_approx}
\eta \approx e^{-\nu_1 I_0^2 c^2 (x - x_1)^4 \tau}.
\end{equation}
This is the functional form in Eq.~1 of the main text used to fit the NV$^-$ population around nodes in Fig.~1(d-e) and Fig.~2(b) of the main text.

\subsection{Resolution as a function of depletion duration}

\begin{figure}[htbp]
\includegraphics[width=0.4\textwidth]{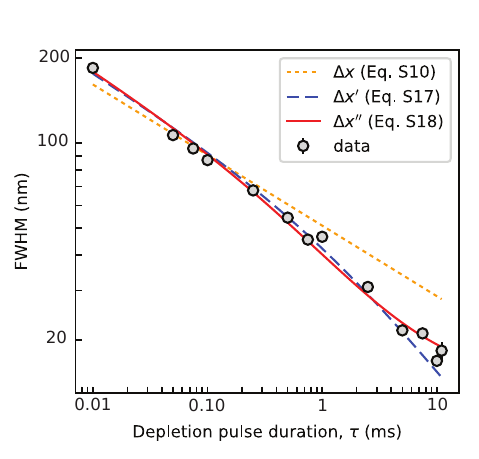}
\centering
\caption{\label{fig:width_scaling} Comparison of fit functions to resolution as a function of depletion duration. Data points (circles) originally plotted in Fig. 2(b) of main text. The orange dotted line is fit to Eq.~\ref{res_scaling_first_approx}. The blue dashed line is fit to Eq.~\ref{background_int_width_scaling} The red solid line is fit to Eq.~\ref{background_int_width_scaling_R}, and is also shown in Fig.~2(b) of the main text.
}
\end{figure}

In this section, we derive the scaling of the SAM resolution with the duration of the depletion pulse, $\tau$. We define the resolution as the FWHM of the peak about one of the nodes ($n_1, n_2, \dots$) of the NV$^-$ population $\eta$. Combined with Eq.~\ref{eta_approx}, this definition can be expressed
\begin{equation}\label{fwhm_def}
\eta = 1/2 = e^{-\nu_1 I_0^2 c^2 (\pm \delta x)^4 \tau}, 
\end{equation}
where $2 \delta x$ is the FWHM about the node at \(x_{1}\). This allows us to examine the relationship between the FWHM and $\tau$ in the argument of the exponential. By rearranging Eq.~\ref{fwhm_def} and defining $\alpha=\ln(2) /(\nu_1 c^2)$, we obtain
\begin{equation}
\alpha = I_0^2 (\pm \delta x)^4 \tau.
\end{equation}
Solving for the FWHM $\Delta x = 2\delta x$
\begin{equation}\label{res_scaling_first_approx}
\Delta x = 2\frac{\alpha^{1/4}}{I_0^{1/2} \tau^{1/4}},
\end{equation}
we see that $\Delta x \propto \tau^{-1/4}$. Also note that this expression gives a scaling of the resolution with laser intensity as well: $ \Delta x \propto I_0^{-1/2}$. Figure~\ref{fig:width_scaling} shows that this first approximation does not accurately describe the experimental data (circles) originally presented in Fig. 2(b) of the main text. We make two additional approximations to match theory with experiment. 

First, we assume that the intensity of the Airy pattern at the node is not zero, but is a small fraction $\epsilon \ll 1$ of the peak intensity $I_0$. This non-zero intensity likely has contributions from several sources, including aberrations associated with focusing through the interface between the air/objective oil and the diamond (which has a higher index of refraction), scattered light, and higher order modes of the depletion beam before the objective. 

Second, we assume that the resolution is limited by the positioning repeatability of our nano-positioner, which we denote $\Delta R$. The first assumption modifies Eq.~\ref{Airy}:
\begin{equation}\label{Airy_mod}
    I(x) = I_0 ( 2 J_1(x) / x)^2 + \epsilon I_0 .
\end{equation}
To the lowest non-zeroth order in \((x-x_1)\), 
\begin{align}
I(x)/I_0 = & ~\epsilon + \frac{2}{x_1^4} [x_1^2J_0(x_1)^2  - 3(x_1^2-4)J_1(x_1)^2 + x_1^2J_2(x_1)^2 - \nonumber \\
& 2x_1J_0(x_1) (4J_1(x_1) + x_1J_2(x_1)) +  
x_1 J_1(x_1)( 8 J_2(x_1) + x_1J_3(x_1))] (x-x_1)^2 + (\dots) \nonumber \\
I(x)/I_0 \approx & ~\epsilon + k (x-x_1)^2 
\end{align}
We write the coefficient for the quadratic term \(k\) and calculate $k \approx 0.08839$ at the first node $x_1 \approx 3.8317$ and $k \approx 0.01464$ at the second node $x_2 \approx 7.0156$.

We see that around the peaks, the intensity is quadratic in $(x-x_1)$ and retains its small constant value $\epsilon$. We put this expression for the intensity into Eq.~\ref{eta}:
\begin{align}
\eta &= e^{-\nu_1 I_0^2 ( (2 J_1(x) / x)^2 + \epsilon)^2\tau}\\
&\approx e^{-\nu_1 I_0^2(k(x-x_1)^2 + \epsilon)^2\tau}\\
&\approx e^{-\nu_1  I_0^2 (k^2(x-x_1)^4 + 2k(x-x_1)^2\epsilon + \mathcal{O}(\epsilon^2))\tau}.
\end{align}

As before, setting \(\delta x' = x-x_1\) and defining
\begin{equation}
    \alpha' = I_0^2 (k^2\delta x'^4 + \epsilon k\delta x'^2)\tau,
\end{equation}
we solve for the FWHM $\Delta x' = 2 \delta x'$, obtaining an expression for the scaling of the FWHM as a function of $\tau$ (or $I_0$)
\begin{equation}\label{background_int_width_scaling}
    \Delta x' = \sqrt{ \frac{2}{k} \left(-\epsilon + \sqrt{\epsilon^2 + 4\alpha'/(I_0^2 \tau)}\right)}.
\end{equation}
Note that we have taken the positive square roots. Equation~\ref{background_int_width_scaling} is fit (blue dashed line) to the data in Fig.~\ref{fig:width_scaling}, which describes the scaling better than Eq.~\ref{res_scaling_first_approx}, however fails to capture the saturation at longer times.

To address the saturation of the resolution at longer depletion durations, we make the second assumption described above, that the positioning repeatability $\Delta R$ is finite and that it places a hard limit on the achievable resolution which is independent of depletion duration. Treating the resolution limit due to the beam profile and the limit due to positioning repeatability as uncorrelated errors, we obtain a total FWHM resolution limit, 
\begin{equation}\label{background_int_width_scaling_R}
\Delta x'' = \sqrt{\Delta x'^2 + \Delta R^2} = \sqrt{\frac{2}{k} \left(-\epsilon + \sqrt{\epsilon^2 + 4\alpha'/(I_0^2 \tau)}\right) + \Delta R^2}.
\end{equation}
Equation~\ref{background_int_width_scaling_R} is plotted in Fig.~\ref{fig:width_scaling} (red solid line). This equation is used to fit experimental data in Fig. 2(b) of the main text, from which we extract values for the non-zero intensity at the node $n_2$ of $\epsilon=0.44(12)\%$ of $I_0$ and the positioning repeatability $\Delta R=14(3)$~nm.

\subsection{Circle fitting routine to determine NV center positions}

\begin{figure}[htbp]
\includegraphics[width=\textwidth]{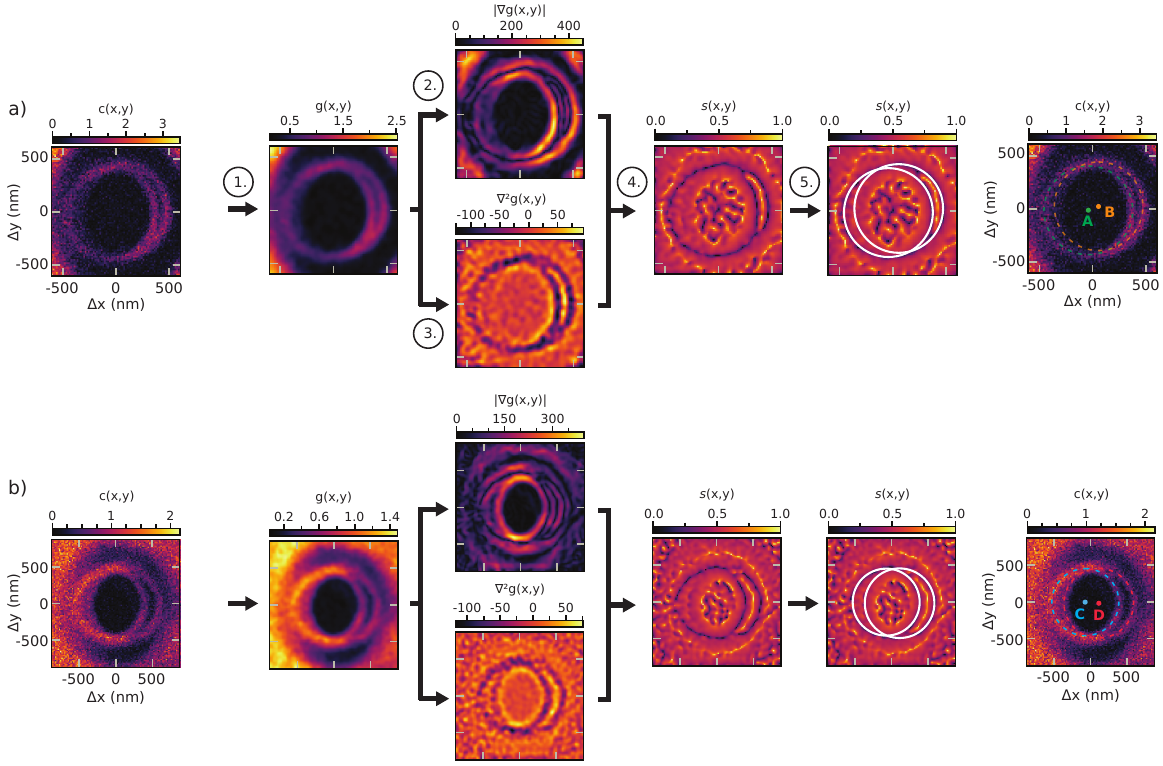}
\centering
\caption{\label{fig:circlefit} Algorithm for finding best fit of a circle to super-resolution rings of (a) NV$_\text{A}$ and NV$_\text{B}$ from Fig. 3 of main text and (b) NV$_\text{C}$ and NV$_\text{D}$ from Fig. 4 of main text. The method starts with the original SAM measurement $c(x, y)$ (left-most panel), and (1) a Gaussian blur is applied. (2) The magnitude of the gradient and (3) the Laplacian are calculated from the blurred image. (4) The ratio of the Laplacian to the magnitude of the gradient is used for the argument of a sigmoid function (Eq.~\ref{sigmoid}) to determine positions of local maxima in the blurred image, corresponding to values of 0 in the image of \(s(x,y)\). Lastly (5) an optimization step finds circles that minimize the line integral of \(s(x,y)\) (Eq.~\ref{cost_function}). This results in the fitted circles plotted to the right (and in the main text) and used to determine the NV center position. 
}
\end{figure}

In order to systematically identify the locations of the NV centers that give rise to the pair of super-resolution rings shown in Figs. 3 and 4 of the main text, we use a custom algorithm designed to identify the circles that best lie along the radial maxima associated with each super-resolution ring. The algorithm relies on three image processing operations, each performed using the open-source computer vision library opencv as implemented in Python \cite{2022opencv}. The kernel size is set to 7 pixels for each operation.

\begin{enumerate}
  \item Gaussian blur: This step removes high frequency noise from the images.
  \item Sobel filter: The Sobel filter approximates a directional derivative along the image. By applying the filter for both the $x$ and $y$ directional derivatives, we obtain the gradient of the image. Small gradient magnitudes indicate areas in the image where the pixel values are changing slowly, which may correspond to radial minima/maxima.
  \item Laplace filter: This filter approximates the Laplacian of the image. The sign of the Laplacian can indicate the local concavity of the image. In combination with the magnitude of the gradient, this can be used to identify radial maxima. 
\end{enumerate}

The algorithm starts by applying a Gaussian blur to the raw image \(c(x,y)\), yielding \(g(x,y)\). The magnitude of the gradient and the Laplacian are then calculated by passing the blurred image through the Sobel filters and the Laplace filter. The Gaussian blur, Sobel filter, and Laplace filter steps are labeled (1), (2), and (3) respectively in Fig.~\ref{fig:circlefit}. The resulting filtered images are next used to assign each pixel in the image a value between 0 and 1, where  0 indicates a flat point with negative Laplacian (e.g. a local maximum) and 1 indicates a flat point with positive Laplacian (e.g. a local minimum). The operation used to achieve this mapping is a sigmoid function whose argument is the quotient of the Laplacian and the magnitude of the gradient:
\begin{align}
    \label{sigmoid} s(x,y) = \left[1+\exp(-\frac{\nabla^{2}g(x,y)}{\abs{\nabla g(x,y)}})\right]^{-1}.
\end{align}
We see that if the magnitude of the gradient at some point is small, then the argument of the exponential will be either large and negative or large and positive depending on the sign of the Laplacian, causing a local maximum in $g(x,y)$ to return 0 and a local minimum to return 1. Pixels with large gradients and small Laplacians will return 0.5. This is shown as step (4) in Fig.~\ref{fig:circlefit}, where the resulting image \(s(x,y)\) reveals clear well-isolated dark arcs associated with the super-resolution radial maxima.

The final step (5) is a simple minimization. For a circle %\(\mathbf{r}_{XYR}(\phi) = (X + R \cos \phi, Y + R \sin \phi)\) 
\(C(X,Y,R)\) of radius \(R\) centered at \((X,Y)\), 
we define a cost function \(S(X,Y,R)\) that returns the average of the line integral of \(s(x,y)\) around the circle:
\begin{align}
    \label{cost_function}
    %S(X,Y,R) &= \frac{1}{2\pi}\oint s(\mathbf{r}_{XYR}) \, d\mathbf{r}_{XYR}.
    S(X,Y,R) &= \frac{1}{2\pi R}\oint_{C(X,Y,R)} s(\mathbf{r}) \, dr.
\end{align}
where \(\mathbf{r}= (x,y)\). 
The uppercase variables \(X\), \(Y\), and \(R\) refer to coordinates in the cost function parameter space while the lowercase variables \(x\) and \(y\) refer to coordinates in the image.
In practice, the average in Eq.~\ref{cost_function} is calculated using 1000 evenly spaced samples around the circle rather than the continuous line integral. 
In order to bypass local minima in the cost function parameter space and avoid excessive biasing from an initial guess, the minimization is conducted by brute force over recursively smaller search regions until the cost of the optimized circle decreases by less than \(0.01\%\) compared to the result from the previous run. 

In order to motivate a definition for the uncertainty of a given circle parameter, we consider an idealized super-resolution ring whose optimum fit circle has parameters \((X',Y',R')\). A point at angle \(\phi\) along the circle has coordinates 
\begin{align}
    \mathbf{r}(X',Y',R'; \phi) = \big(X' + R' \cos \phi, \, Y' + R' \sin \phi\big),
\end{align}
which we abbreviate with \(\mathbf{r}'(\phi)\). 
We assume that the value of \(s(x,y)\) 
changes linearly at rate $A$ 
along the radius of the circle for small deviations \(\delta \ll R'\) from the optimum radius:
\begin{align}\label{idealized_ring}
    %s\big(\mathbf{r}_{X'Y'(R'+\delta)}\big) = A\abs{\delta}.
    %s\big(X' + (R' + \delta) \cos \phi, \, Y' + (R' + \delta) \sin \phi\big) = A\abs{\delta}
    s\big(\mathbf{r}(X', Y', R'+\delta; \phi)\big) = A\abs{\delta}.
\end{align} 
We now introduce the functions 
\begin{gather}
    f(x,y)=0.5-s(x,y),\\
    F(X,Y,R)=0.5-S(X,Y,R),
\end{gather}
which convert minima of \(s(x,y)\) and \(S(X,Y,R)\) into peaks with heights set in relation to 0.5, the expected value of \(s(x,y)\) for a random point. For the idealized ring, we define the uncertainty for the $i$th circle parameter as the half width at half max of the peak in \(f(x,y)\) along a 
normal of the circle which is parallel to the direction associated with a change in the $i$th circle parameter.
Intuitively, this quantifies how much we can shift the optimum circle before it becomes obvious that the shifted circle no longer provides a good fit to the ring. More formally, the uncertainty \(\sigma_{i}\) for the \(i\)th circle parameter is defined as half the difference between the values of \(\Delta\) that satisfy
\begin{align}
    %f(\mathbf{r}^{*} + \Delta \mathbf{i}(\mathbf{r}^{*})) = \frac{1}{2}f(\mathbf{r}^{*}),
    f\big(\mathbf{r}'(\phi^{*}) + \Delta \mathbf{i}(\phi^{*})\big) = \frac{1}{2}f\big(\mathbf{r}'(\phi^{*})\big),
\end{align}
where \(\mathbf{i}(\phi)\) is the unit vector associated with a change in the \(i\)th circle parameter at angle \(\phi\) along the circle 
%\(\mathbf{r}'\) 
and \(\phi^{*}\) is an angle 
%\(\mathbf{r}^{*}\) is a point along \(\mathbf{r}'\) 
that solves \((d\mathbf{r}'(\phi) / d\phi) \cdot \mathbf{i}(\phi) = 0\).
We now wish to generalize this equation to take into account the entire circle rather than just a select point. As an initial step, we use Eq.~\ref{idealized_ring} to find a relationship between \(\sigma_{i}\) and the function \(F\) evaluated for the optimum circle. We see that
%With the abbreviations \(\mathbf{r}^{*} = \mathbf{r}'(\phi^{*})\) and \(\mathbf{i}^{*} = \mathbf{i}(\phi^{*})\),
\begin{align}
    %f(\mathbf{r}^{*}) - f(\mathbf{r}^{*} + \delta \mathbf{i}^{*}) = A \abs{\delta}
    f\big(\mathbf{r}'(\phi^{*})\big) - f\big(\mathbf{r}'(\phi^{*}) + \delta \mathbf{i}(\phi^{*})\big) = A \abs{\delta}
\end{align}
and so
\begin{align}
    \sigma_{i} &= \frac{1}{2A}f\big(\mathbf{r}'(\phi^{*})\big) \nonumber\\
    \label{fwhm_def_ring} &= \frac{1}{2A}F(X',Y',R').
\end{align}
We next calculate the value of \(F\) for an optimum circle with its \(i\)th parameter shifted by \(\Delta\). 
Denoting this quantity 
\begin{align}
    F'_{i}(\Delta) %&= \frac{1}{2} - \frac{1}{2\pi}\oint s\big(\mathbf{r}' + \Delta \mathbf{i}(\mathbf{r}')\big) \, d\mathbf{r}'
    &= \frac{1}{2} - \frac{1}{2\pi}\int s\big(\mathbf{r}'(\phi) + \Delta \mathbf{i}(\phi)\big) \, d\phi
\end{align}
and with \(F' = F(X',Y',R')\), we first calculate the differential effect of the shift:
\begin{align}
    F' - F'_{i}(\Delta) 
    %&= \frac{1}{2\pi}\oint \Big[s\big(\mathbf{r}' + \Delta \mathbf{i}(\mathbf{r}')\big) - s\big(\mathbf{r}'\big)\Big] \, d\mathbf{r}' \nonumber\\
    &= \frac{1}{2\pi}\int \Big[s\big(\mathbf{r}'(\phi) + \Delta \mathbf{i}(\phi)\big) - s\big(\mathbf{r}'(\phi)\big)\Big] \, d\phi \nonumber\\
    \label{double_int} &= \frac{1}{2\pi}\int \bigg[\int_{0}^{\Delta}\nabla_{\mathbf{i}(\phi)}s\big(\mathbf{r}'(\phi) + t \mathbf{i}(\phi)\big) \, dt \bigg] \, d\phi.
\end{align}
where 
%we have %parameterized the integral around the circle by the angle \(\phi\), 
%introduced the dummy variable \(t\). 
%and made use of the abbreviation \(\mathbf{i}(\phi)=\mathbf{i}(\mathbf{r}'(\phi))\). 
%We use 
\(\nabla_{\mathbf{i}(\phi)}s(x,y)\) denotes 
%to denote 
the directional derivative of \(s\) along \(\mathbf{i}(\phi)\) at \((x,y)\). %the unit vector for the \(i\)th circle parameter. 
For \(\Delta \ll R\), we use Eq.~\ref{idealized_ring} to calculate
\begin{align}
    \int_{0}^{\Delta}\nabla_{\mathbf{i}(\phi)}s\big(\mathbf{r}'(\phi) + t \mathbf{i}(\phi)\big) \, dt &\approx \Delta \, \abs{\nabla_{\mathbf{i}(\phi)}s\big(\mathbf{r}'(\phi)\big)} \nonumber\\
    &= \Delta An_{i}(\phi) \label{single_int_eval} 
\end{align}
where \(n_{i}(\phi) = \abs{\mathbf{i}(\phi) \cdot \mathbf{n}(\phi)}\) and \(\mathbf{n}(\phi)\) is the unit normal of the circle at angle \(\phi\).
Combining Eqs. \ref{single_int_eval} and \ref{double_int}, we see that 
\begin{align}
    F' - F'_{i}(\Delta) &= \frac{\Delta A}{2\pi}\int n_{i}(\phi) \, d\phi.
\end{align}
or
\begin{align}
    F'_{i}(\Delta) &= F' - \frac{\Delta A}{2\pi}\int n_{i}(\phi) \, d\phi.
\end{align}
Using Eq.~\ref{fwhm_def_ring} to evaluate this expression at \(\Delta = \sigma_{i}\), we find that \(\sigma_{i}\) is also half the difference between the values of \(\Delta\) that satisfy
\begin{align}
    F'_{i}(\Delta) &= F' \cdot \left(1-\frac{1}{4\pi}\int n_{i}(\phi) \, d\phi\right),
\end{align}
which is the basis for our final definition of the uncertainty. 
Specifically, we define the uncertainty \(\sigma_{i}\) for the \(i\)th parameter as the half width at \(c_{i}\) of the maximum about the peak in \(F'_{i}(\Delta)\) at \(\Delta=0\). The factors \(c_{i}\) are defined
\begin{align}
    c_{i} = 1-\frac{1}{4\pi}\int n_{i}(\phi) \, d\phi.
\end{align}
Precisely, \(c_{X} = c_{Y} = 1-\pi^{-1} \approx 0.68\) and \(c_{R} = 1/2\).

The results of the fitting and uncertainty calculations are presented in Table~\ref{table:ring_fitting} in pixels. The right-most panels in Fig.~\ref{fig:circlefit} show the resultant circles that describe the super-resolution rings. We calculate that NV$_{\text{A}}$ and NV$_{\text{B}}$ (displayed in Fig.~3 of the main text) are separated by 113(26)~nm and NV$_{\text{C}}$ and NV$_{\text{D}}$ (displayed in Fig.~4 of the main text) are separated by 181(26)~nm.
% Fig. 3: 7.5(1.7) pixels, 15.225~nm / pixel
% Fig 4: 10.5(1.7) pixels, 17.4~nm / pixel

The circle fitting routine described in this section is also used to produce the super-resolution reconstruction images  shown in Figs.~3 and 4 of the main text. 
%Lastly, Figs.~3 and 4 of the main text show reconstructed positions of the NV centers based on this fitting routine. 
Specifically, the pixel value at each point \((x,y)\) in a reconstruction image is the cost function for a circle with center \((X=x, Y=y)\) and optimized radius \(R' = \text{argmin}_{R}\big(S(X,Y,R)|_{X=x,Y=y}\big)\). The domain of the radius optimization is limited to a small range centered on the average of the radii of the two superresolution rings in each image. 
%Specifically, for each point \((x,y)\) in the reconstruction image, we calculate the cost function Eq.~\ref{cost_function} for circles with centers $(X=x, Y=y)$, and select the minimum cost function over a range of values for the circle's radius $R$. 
%The color maps in Figs.~3 and 4 of the main text plot the minimized cost function value, 
Pixels with lower values (lighter colors in the images) correspond to points where an NV center is more likely to be located.

\begin{table}
\centering
\begin{tabular}{ |c|c|c|c|c| } 
 \hline
  & X center (nm) & Y center (nm) & Radius (nm) \\ 
 \hline
NV$_{\text{A}}$ & -47(21) & -26(23) & 418(18)\\
NV$_{\text{B}}$ & 59(14) & 14(20) & 417(15)\\
NV$_{\text{C}}$ & -72(17) & -17(24) & 450(22)\\
NV$_{\text{D}}$ & 109(19) & -21(24) & 471(21)\\
 \hline
\end{tabular}
\caption{Center coordinates and radii for the super-resolution rings shown in Fig.~\ref{fig:circlefit}. The super-resolution rings of NV$_{\text{A}}$ and NV$_{\text{B}}$ are shown in Fig.~3 of the main text and Fig.~\ref{fig:circlefit}(a). The super-resolution rings of NV$_{\text{C}}$ and NV$_{\text{D}}$ are shown in Fig.~4 of the main text and Fig.~\ref{fig:circlefit}(b).  The parameters are calculated using the algorithm described here and depicted in Fig.~\ref{fig:circlefit}. }
\label{table:ring_fitting}
\end{table}

\subsection{Simulating SAM measurements}
\label{SimSam}

In a previous section we examined the intensity profile of the depletion beam for a near-perfect Airy disk near the Airy nodes. The resulting model does a good job of describing the scaling of the resolution with depletion duration (Fig.~\ref{fig:width_scaling}), but it fails to describe the SAM measurement far from the Airy nodes. In this section we describe two additional methods of simulating SAM measurements more realistically, which were pursued with the intention of accurately reproducing experimental data. The methods model the system using (1) the angular spectrum representation technique and (2) the commercial software Lumerical.

\begin{figure}[htbp]
\includegraphics[width=\textwidth]{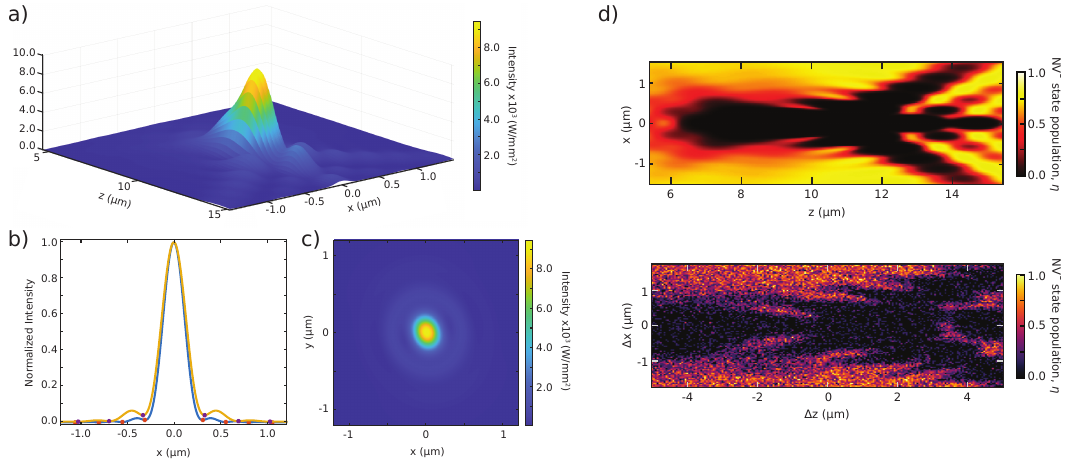}
\centering
\caption{\label{fig:simulation_1} ASR simulations of SAM measurements. (a) Surface plot of simulated laser intensity in the $xz$-plane demonstrating aberrations associated with oil-diamond interface. (b) Simulated laser intensity taken along x direction at the focus within the diamond $z=$ 10.46~\textmu m (orange curve) and at the focus %$z =$~10.5 \textmu m 
with the diamond removed (blue curve). Local minima of the intensity curves are marked with purple and orange dots respectively. (c) Simulated laser intensity plotted in $xy$-plane at a depth in the diamond $z=$ 10.46~\textmu m. (d) Comparison of simulated (top) and experimental (bottom) SAM measurement in $xz$-plane. The simulation plot is centered around the focus at $z=10.46$~\textmu m.
}
\end{figure}

For the first approach we use the angular spectrum representation (ASR) technique (adapted from Ref.\cite{novotny2012nano_optics}) which  is a useful tool for modeling the electric field in systems involving high-NA objectives. The ASR technique allows the full electric vector field in any given plane along an optical axis to be calculated based on the field at an initial plane elsewhere along the same axis. Optical elements are then modeled by including their effects on the spatial frequencies of the field. We model the objective as an aplanatic converging lens with NA $ = 1.3$ and a focal length of 160~\textmu m, which is the working distance of the oil objective used in the experiments. The initial field is a collimated Gaussian beam of wavelength 638~nm overfilling the lens aperture by a factor of 5. The optical power of the beam used in the simulations is 30~mW. We model the oil-diamond interface as a plane normal to the optical axis with $n_{\text{oil}} = 1.479$ ($ n_{\text{d}} = 2.4$) for the refractive index of the oil (diamond). % located at $z_0$.
The diamond surface is 154.7~\textmu m from the lens. The high refractive index of diamond effectively increases the focal length of the lens such that the beam focus (i.e. the point of highest intensity) occurs at 10.46~\textmu m below the diamond surface.

With the ASR method, we obtain the intensity of the beam at an arbitrary position inside the diamond. In the following descriptions and plots of the theory results, we assume a Cartesian coordinate system where the \(xy\)-plane at \(z=0\) coincides with the oil-diamond interface and positive \(z\) coordinates indicate depth in the diamond. The polarization vector of the input Gaussian beam used in the ASR method is $((1+2i)\mathbf{e}_x + 4\mathbf{e}_y)/\sqrt{21}$, which is mainly linearly polarized along the $y$ direction (see next section for more information). The intensity in the \(xz\)-plane is plotted in Fig.~\ref{fig:simulation_1}(a) for a beam with optical axis passing through the origin of the coordinate system. Figure~\ref{fig:simulation_1}(b) plots the intensity at the beam's focus at z = 10.46~\textmu m to highlight some of the aberrations introduced by the objective oil-diamond interface, such as shifts in the positions of the intensity minima and increased intensities at the minima. Figure~\ref{fig:simulation_1}(c) plots the intensity in the $xy$-plane and shows dark rings similar to the nodes of a pure Airy pattern. However, the intensity in these rings is not exactly zero, and so will eventually deplete an emitter under a sufficiently long depletion pulse in a SAM measurement. As mentioned in the main text, we believe this is the main reason that the bright rings of the SAM measurement eventually disappear for long depletion pulse durations.

The SAM measurement is simulated with Eq.~\ref{NV-_rate}:
\begin{equation}\label{sim_NV-_rate}
    \eta(x,y,z) = e^{-\nu_1 I(x, y, z)^2 \tau},
\end{equation}
which uses the laser intensity calculated form the ASR model \(I(x,y,z)\) and the constant \(\nu_1\) fixed to \(\nu_1 = 2.88\times10^{-4}~\text{mm}^4/(\text{ms}\cdot\text{W}^2)\). Equation~\ref{sim_NV-_rate} effectively simulates the NV$^-$ population, $\eta$, of an NV center at position $(x, y, z)$ for a beam centered at ($x=0$ \textmu m, $y=0$ \textmu m) and focused at $z=10.46$~\textmu m below the diamond surface.
Figure~\ref{fig:simulation_1}(d) (top) plots the simulated SAM measurement in the $xz$-plane based on the intensity in Fig.~\ref{fig:simulation_1}(a). 
The bottom panel plots the analogous experimental SAM measurement, which qualitatively agrees well with the simulation. These plots reveal how the width and quality of the dark rings of the SAM measurement strongly depend on the axial as well as radial offsets of the depletion beam with respect to the target emitter. We also note that the SAM technique does not significantly increase the resolution in $z$, which is common with many other super-resolution techniques such as STED and GSD \cite{hell2015sr_review, schermelleh2019sr_review}

In addition to the ASR model, we use commercially available software (Lumerical, Inc.), which solves Maxwell's equations on a rectangular grid within the finite difference time-domain (FDTD) scheme \cite{hagness2005computational} to simulate the beam propagation through a lens of NA 1.3 positioned 40 nm from a diamond surface. The ``distance from focus" parameter for the lens in the software is set to 5.5~\textmu m, which, because of the high refractive index of diamond, places the focus 10.59~\textmu m within the diamond. The objective oil and diamond refractive indices are the same as noted above.  In this simulation, the dimensions of the diamond are 25~\textmu m $\times$ 25~\textmu m $\times$ 21~\textmu m and the origin of the coordinate system is at the center of the diamond's square face. In the software, the thin lens approximation is applied, and the option to fill the lens is chosen. Lumerical calculates the electric field magnitude, $|E(x,y,z)|^2$, at position $(x,y,z)$. For simulating SAM measurement, $|E(x,y,z)|^2$ is normalized to the maximum electric field magnitude in the diamond (occurring at the focus): $|E(x,y,z)|^2 / E_{\text{max}}^2$. This normalized intensity is used in Eq.~\ref{sim_NV-_rate}:
\begin{equation}
    \eta(x,y,z) = e^{-\nu_1 I_0^2 \big(|E(x,y,z)|^2 / E_{\text{max}}^2\big)^2 \tau },
\end{equation}
where the combined value of $\nu_1 I_0^2$ is fixed to $\nu_1 I_0^2 =2 \times 10^9$~ms$^{-1}$.

\begin{figure}[htbp]
\includegraphics[width=\textwidth]{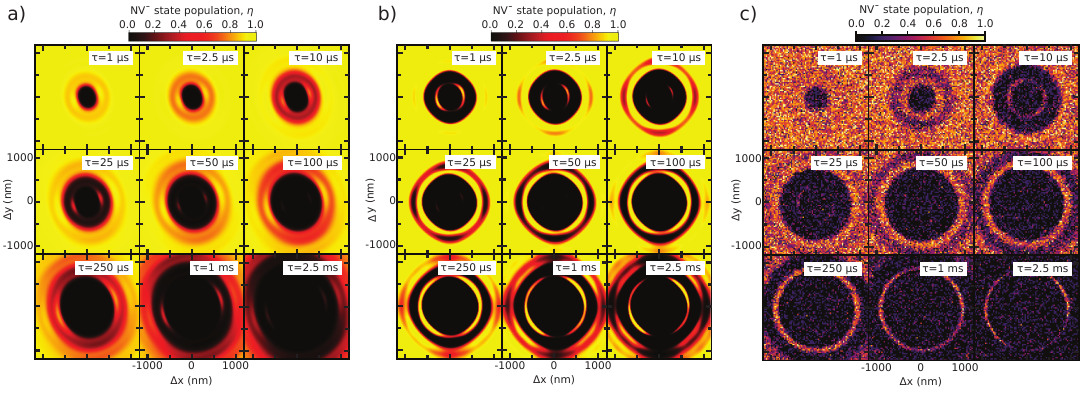}
\centering
\caption{\label{fig:simulation_comps} (a) Simulated SAM measurements for various depletion pulse durations $\tau$ using laser intensity from the ASR model at a depth of 10.46~\textmu m in the diamond. 
(b) Simulated SAM measurements for various $\tau$ using laser intensity from Lumerical at a depth of 10.59~\textmu m in the diamond. (c) Extended experimental data from Fig. 2 of the main text. NV center is estimated to be about 10~\textmu m deep in the diamond.
}
\end{figure}

SAM measurements for different depletion durations \(\tau\) are simulated using the laser intensities calculated from the two models. The results are shown alongside experimental measurements in Fig.~\ref{fig:simulation_comps}. The simulations shows good qualitative agreement with experiment, and replicate the primary features seen in the experimental measurements. Specifically, the simulations capture the disappearing of the rings at longer depletion durations, which is mainly due to the aberrations induced by the diamond interface. We believe the achievable resolution of the SAM technique can be improved by reduced these aberrations. In addition, the simulations capture the radial asymmetry of the bright rings. In the following section, we discuss this effect and attribute it to the linear polarization of the depletion beam used in the experiments. We also observe substantial differences between simulation and experiment, such as disagreement in the diameter of the super-resolution rings and the values of $\tau$ at which the rings disappear. More work is needed to show quantitative agreement between theory and experiment, which might be achieved using a more realistic model of the objective.  

\subsection{Polarization dependence of SAM measurement}

\begin{figure}[htbp]
\includegraphics[width=\textwidth]{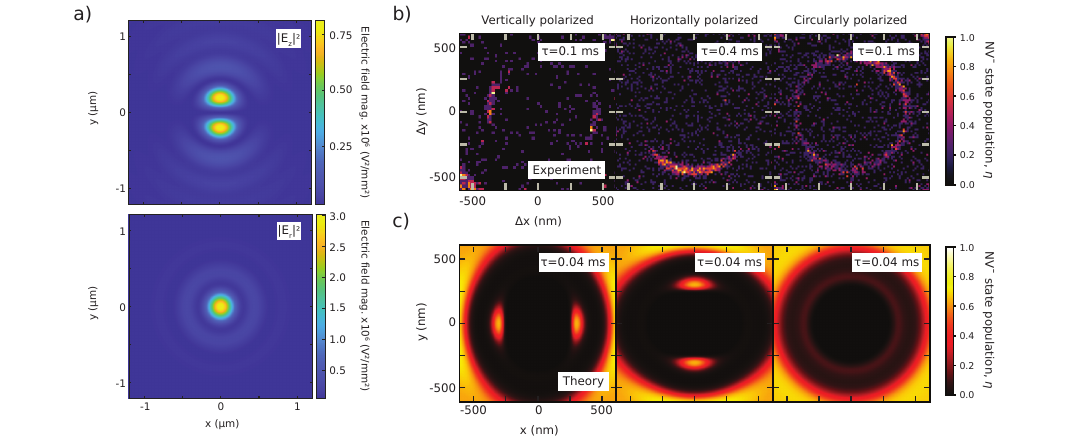}
\centering
\caption{\label{fig:polarization} Optical polarization effects on SAM measurement. (a) Electric field magnitude in $z$, $|E_z|^2$ (top) and in radial direction, $|E_r|^2$  (bottom) calculated using ASR technique for vertically polarized 638~nm light focused into diamond with a 1.3 NA oil objective. (b) Experimental measurement of the $n_1$ ring in response to manipulating the polarization of the depletion beam. (c) Simulated SAM measurements using ASR technique with different polarization of depletion beam.
}
\end{figure}

The polarization of light passing through a high-NA lens can result in significant breaking of radial symmetry in the intensity distribution about the optical axis \cite{novotny2012nano_optics}.
We observe this effect in both experiment and simulation. With linearly polarized light, the $z$-component of the electric field, $E_z$, can be significantly greater than zero along the axis of polarization in any given plane. This is shown in Fig.~\ref{fig:polarization}(a), which plots the magnitude of $|E_z|^2$ (top) and the magnitude of the radial electric field $|E_r|^2 = (|E_{x}|^{2}+|E_{y}|^{2})$ (bottom) for vertically polarized light at same plane plotted in Fig.~\ref{fig:simulation_1}c. These plots show that the $E_z$ component contributes significant intensity along the direction of polarization, including at positions where the nodes in the $E_r$ profile are near zero. This has a noticeable impact on the SAM measurements, where the quality of the super-resolution rings depends on the intensity of the nodes being near zero. This effect causes the super-resolution rings to vanish faster along the polarization axis as compared to the axis perpendicular to the polarization. 

We quantify the effect of the depletion light polarization on the SAM measurements in experiment (Figure~\ref{fig:polarization}(b)) and in ASR simulations (Figure~\ref{fig:polarization}(c)). For the experimental measurements, the beam polarization was rotated with wave plates before entering the objective. Both theory and experiment exhibit the same behavior, where bright arcs form along the direction perpendicular to the polarization because the optical intensity minima along this direction are closer to zero versus the direction parallel to the polarization. Simulations show an ellipticity in the pattern along the direction of polarization which is not seen in experiments. For circular polarization, the
radial symmetry of the ring is restored, however at the cost of non-zero intensity along the whole ring because the $E_z$ component is averaged around the ring.

For data shown in the main text, the 638~nm light was found to be linearly polarized at a small angle from the vertical by the dichroic mirrors used in our experiments. 
We note that in experiment, the orientation of the bright arcs appears to change between the $n_1$ and $n_2$ node, as can be seen in Fig.~\ref{fig:simulation_comps}(c) panels for $\tau = 10$~\textmu s and $\tau = 2.5$~ms. For the simulations shown in Fig.~\ref{fig:simulation_comps}(a) and (b), the polarization of the beam was adjusted to best match the $n_2$ node in experiment (see $\tau = 2.5$~ms in Fig.~\ref{fig:simulation_comps}(c)). 
For the simulations using the ASR model shown in Fig.~\ref{fig:simulation_comps}(a), the polarization vector of the input electric field was chosen to be $((1+2i)\mathbf{e}_x + 4\mathbf{e}_y)/\sqrt{21}$, which is a combination of nearly vertical linear polarization and circular polarization. For the Lumerical simulations shown in Fig.~\ref{fig:simulation_comps}(b), the polarization was taken to be linear and oriented $20 \degree$ counterclockwise from vertical.
\bibliography{MainReferences}